\documentstyle{article}
\begin{document}

\begin{center}
{\Large
         Method of estimating distances
         to X-ray pulsars and their magnetic fields
}
\vskip 0.5cm
{\bf S.B. Popov}\\

Sternberg Astronomical Institute\\

\vskip 0.5cm

119899, Russia, Moscow,\\
Universitetski pr. 13\\
e-mail: polar@xray.sai.msu.su\\

\vskip 0.5cm

{\bf    Abstract}

\end{center}

 We suggest a new method of estimating distances to X-ray pulsars
and their magnetic fields.
Using observations of fluxes and period variations in the model
of disk accretion one can estimate the magnetic momentum of a neutron star
and the distance to X-ray pulsar.

 As an illustration the 
method is applied to the system GROJ1008-57. Estimates of the
distance: approximately 6 kpc, and the magnetic momentum: approximately
$4\cdot10^{31}\, E\cdot cm^3$, are obtained.


\section{Introduction}

 Neutron stars (NS) can appear as isolated objects (radiopulsars, old isolated
accreting NS, etc.) and as X-ray sources in close binary systems.
The most prominent of the last ones are X-ray pulsars, where important
parameters of NS can be determined.

 Now  more than 40 X-ray pulsars are known 
(see, for example, Bildsten et al., 1997). Observations of optical
counterparts give an opportunity to obtain distances to these objects with
high precision, and with hyroline detections one can obtain the value of
magnetic field of a NS. But lines are not detected in all sources of that
type (partly because they can lay out of the range of necessary
spectral sensitivity of devices, when field are high), 
and magnetic field can be estimated
from period measurements (see Lipunov, 1982, 1992). Precise distance
measurements usually are not available immediately after X-ray discovery
(especially, if error boxes, as for example in the BATSE case, are large).
So, methods of their determination basing only on X-ray observations can be
useful.

 Here we propose a simple method to determine magnetic field and
distance to X-ray pulsar using only X-ray 
flux and period variations measurements.   


\section{Method}

 In Lipunov (1982) it was proposed to use maximum spin-up and spin-down
values to obtain limits on the 
magnetic momentum of X-ray pulsars in disk or wind
models, using known values of luminosity (method, based on maximum spin-down,
is very insensitive to uncertainties in luminosity and produces better
results).

 In this short
note we propose a rough simple method to determine magnetic field
without known distance and to determine distance itself. 
The method is based on several measurements
of period derivative, $\dot p$, and X-ray pulsar's flux, $f$.
Fitting two parameters: distance, $d$, and magnetic momentum, $\mu$,
one can obtain good correspondence with the observed $\dot p$ and $f$,
and that way produce good estimates of distance and magnetic field.

 Here we consider only disk accretion. In that case one can write
(see Lipunov, 1982, 1992):

\begin{equation}
\frac {dI\omega}{dt} = \dot M \left( GM\epsilon R_A \right)^{1/2}-
k_t\frac{\mu^2}{R_c^3},
\end{equation}

where $\omega$ -- spin frequency of a NS, $M$ -- its mass, $I$-- its moment
of inertia, $R_A$ - Alfven radius, $R_c$ -- corotation radius.
We use the following values: $\epsilon=0.45$, $k_t=1/3$ (see Lipunov, 1992).
The first term on the right side
represents acceleration of a NS from an accretion disk,
and the second term represents deceleration. The form of the deceleration
term is general, only typical radius of interaction should be changed.
It is equal to 
$R_c$ for accretors, $R_l$ - light cylinder radius for ejectors,
and $R_A$ for propellers (see the details in Lipunov 1992).

Lets rewrite  eq. (1) in terms of period and its derivative:

\begin{equation}
\dot p= \frac {4 \pi ^2 \mu ^2}{3\,G\,I\,M} - (0.45)^{1/2} 2^{-1/14}\frac
{\mu^{2/7}}{I} \left(GM\right)^{-3/7} \left[p^{7/3}L\right]^{6/7}R^{6/7},
\end{equation}

where $L=4\pi d^2 \cdot f$ -- luminosity, $f$ -- observed flux.

So, in eq. (2) we know all parameters ($I$, $M$, $R$ etc.) 
except $\mu$ and $d$.
Fitting observed points with them we can obtain estimates of $\mu$ and $d$.
If $\mu$ is known, one can immediately obtain $d$ from eq. (2) even from 
one determination of $\dot p$ (in that case it is better to use spin-down
value). Uncertainties mainly depend on applicability of
that simple model.


\section{Illustration of the method}

 To illustrate the method, we apply it to the X-ray pulsar 
GRO J1008-57, discovered by BATSE (Bildsten et al., 1997). It is a 
$93.5 \, s$ X-ray pulsar, with the flux about $10^{-9} erg\, cm^{-2}
s^{-1}$. A 33 day outburst was observed by BATSE in August 1993.
The source was also 
observed by EXOSAT (Macomb et al., 1994) and ASCA (Day et al., 1995).
ROSAT made possible to localize the source with high precision
(Petre \& Gehrels, 1994), and it was
identified with a Be-system (Coe et al., 1994)
with $\sim 135^d$ orbital period (Shrader et al. 1999).
We use here only 1993 outburst, described in Bildsten et al. (1997).

 The authors in Bildsten et al. (1997) show flux and frequency history
of the source with 1 day integration. In the maximum of the burst errors
are rather small, and we neglect them. Points with large errors were not
used.

 We used standard values of NS parameters: $I=10^{45}\, g\, cm^2$, moment of
inertia; $R=10\, km$, NS radius; $M=1.4M_{\odot} $, NS mass.

 On figures 1-2 we show observations (as black dots)
and theoretical curves (in disk model,
see Shrader et al. 1999, who proposed a disk formation during the outbursts,
in contrast with Macomb et al. (1994), who proposed wind accretion)
on the plane $\dot p$ -- $p^{7/3} f$, where $f$ -- observed flux (logarithms
of these quantities are shown).
Curves were plotted for different values of the source distance, $d$,
and NS magnetic momentum, $\mu$. 

 The best fit (both for spin-up and spin-down)
 gives $d\approx 5.8\, kpc$ and $\mu\approx 37.6\cdot 10^{30}\, E\cdot cm^3$.
It is shown on both figures. The distance is in correspondence with
the value in Shrader et al. (1999), and such field value is not unusual
for NS in general and for X-ray pulsars in particular (see, for example,
Lipunov, 1992 and Bildsten et al., 1997). Tests on some other X-ray pulsars
with know distances and magnetic fields also showed good results.


\section{Discussion and conclusions}

 The method is only approximate and depends on several assumptions (disk
accretion, specified values of $M, I, R$, etc.). 
Estimates of $\mu$, for example, can be only in rough correspondence with
observations of magnetic field $B$, if standard value of the NS radius,
$R=10\, km$ is used (see, for example,
the case of Her X-1 in Lipunov 1992). Non-standard
values of $I$ and $M$ can also make the picture more complicated.

 The method can be, in principal, generalized for applications to
wind-accreting systems, and to disk-accreting 
systems with complicated time behavior (when, for example, 
$\dot p$ changes appear with nearly
constant flux, or even when $\dot p$ changes are uncorrelated with flux
variations).

If one uses maximum spin-up, 
or maximum spin-down values to evaluate parameters of the pulsar,
then one can obtain values different from the best fit
(they are shown on the figures): $d\approx 8 \, kpc$, 
$\mu\approx 37.6\cdot 10^{30}\, E\cdot cm^3$
for maximum spin-up, and two values
for maximum
spin-down: $d\approx 4 \, kpc$, 
$\mu\approx 37.6\cdot 10^{30}\, E\cdot cm^3$ and the one close to our
best fit (two similar values of maximum spin-down were observed
for different fluxes, but we mark, that formally maximum spin-down
corresponds to the values, which are close to our best fit). 
It can be used as an estimate of the errors of our method:
accuracy is about the factor of 2 in distance, and about the same value in
magnetic field, as can be seen from the figures.

In some very uncertain situations, for example, when only X-ray observations
without precision localization are available, 
our method can give (basing on 
several observational points, not one!, as, for example,
 in the case of maximum spin-down
determination of magnetic momentum), rough, but useful estimates
of important parameters: distance and magnetic momentum.


{\bf Acknowledgments}

It is a pleasure to thank prof. V.M. Lipunov for numerous discussions
and suggestions and drs. I.E. Panchenko, M.E. Prokhorov and K.A. Postnov for
useful comments.
The work was supported by the RFBR (98-02-16801) and
the INTAS (96-0315) grants.

\pagebreak
\vskip 1cm  

{
\Large{Figure captions}
}

\vskip 1cm

Figure 1. 

Dependence of period derivative, $\dot p$, on the parameter $p^{7/3}f$, 
$f$-- observed flux. Both axis are in logarithmic scale.
Observations (Bildsten et al., 1997) are shown with black dots.
Five curves are plotted for disk accretion for different values of
distance to the pulsar and NS magnetic momentum. Solid curve: $d=4 \, kpc$,
$\mu=37.6\cdot 10^{30}\, E\cdot cm^3$. 
Dashed curve: $d=8 \, kpc$, $\mu=37.6\cdot 10^{30}\, E\cdot cm^3$.
Long dashed curve: $d=5.8 \, kpc$, $\mu=10\cdot 10^{30}\, E\cdot cm^3$.
Dot-dashed curve: $d=5.8 \, kpc$, $\mu=45\cdot 10^{30}\, E\cdot cm^3$. 
Dotted curve (the best fit): $d=5.8 \, kpc$, $\mu=37.6\cdot 10^{30}\, E\cdot
cm^3$.

\vskip 0.5cm

Figure 2. 

Dependence of period derivative, $\dot p $, on the parameter $p^{7/3}f$, 
$f$-- observed flux. Both are axis in logarithmic scale.
Observations (Bildsten et al., 1997) are shown with black dots.
Five curves are plotted for disk accretion for different values of
distance to the pulsar and NS magnetic momentum. Solid curve: $d=4 \, kpc$,
$\mu=10\cdot 10^{30}\, E\cdot cm^3$. 
Dashed curve: $d=8 \, kpc$, $\mu=10\cdot 10^{30}\, E\cdot cm^3$.
Long dashed curve: $d=8 \, kpc$, $\mu=45\cdot 10^{30}\, E\cdot cm^3$.
Dot-dashed curve (the best fit): $d=5.8 \, kpc$, $\mu=37.6\cdot 10^{30}\,
E\cdot cm^3$. 
Dotted curve: $d=4 \, kpc$, $\mu=45\cdot 10^{30}\, E\cdot cm^3$.

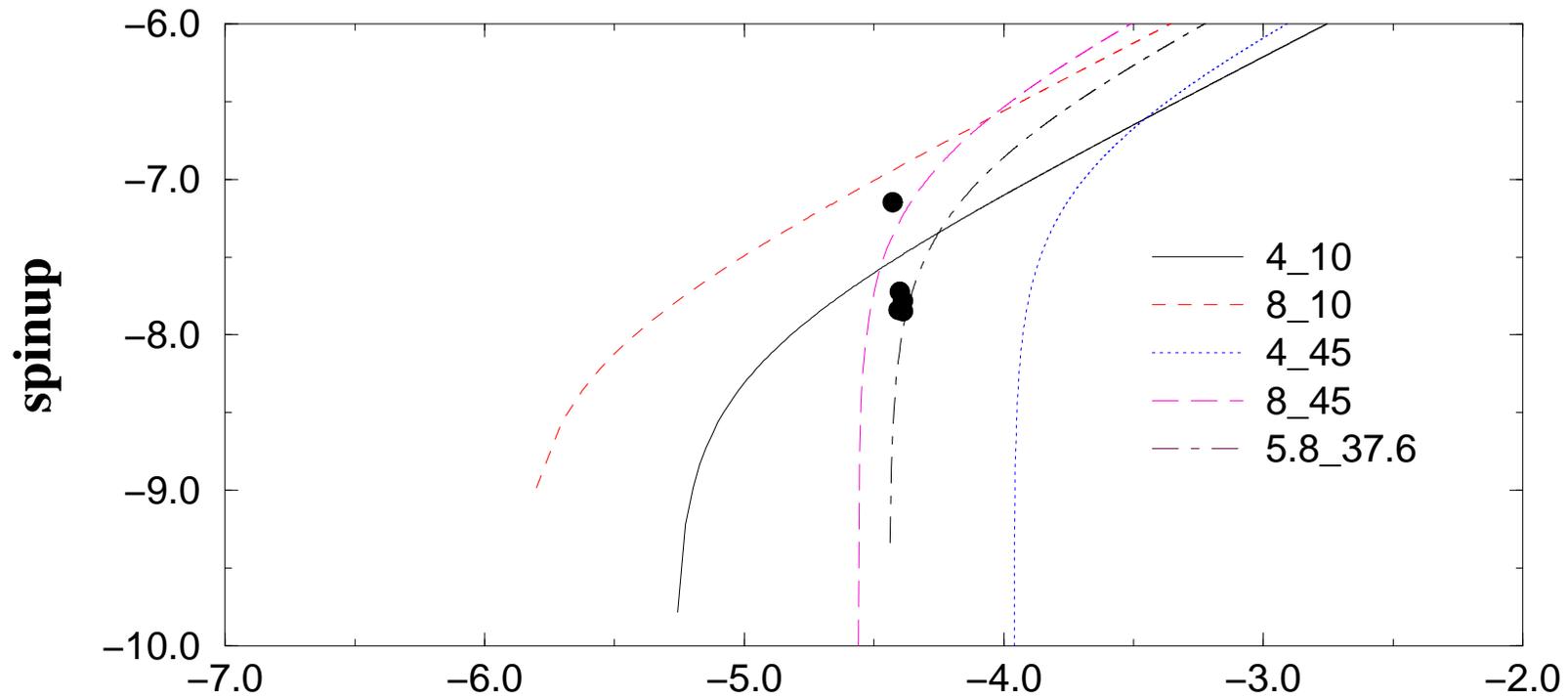
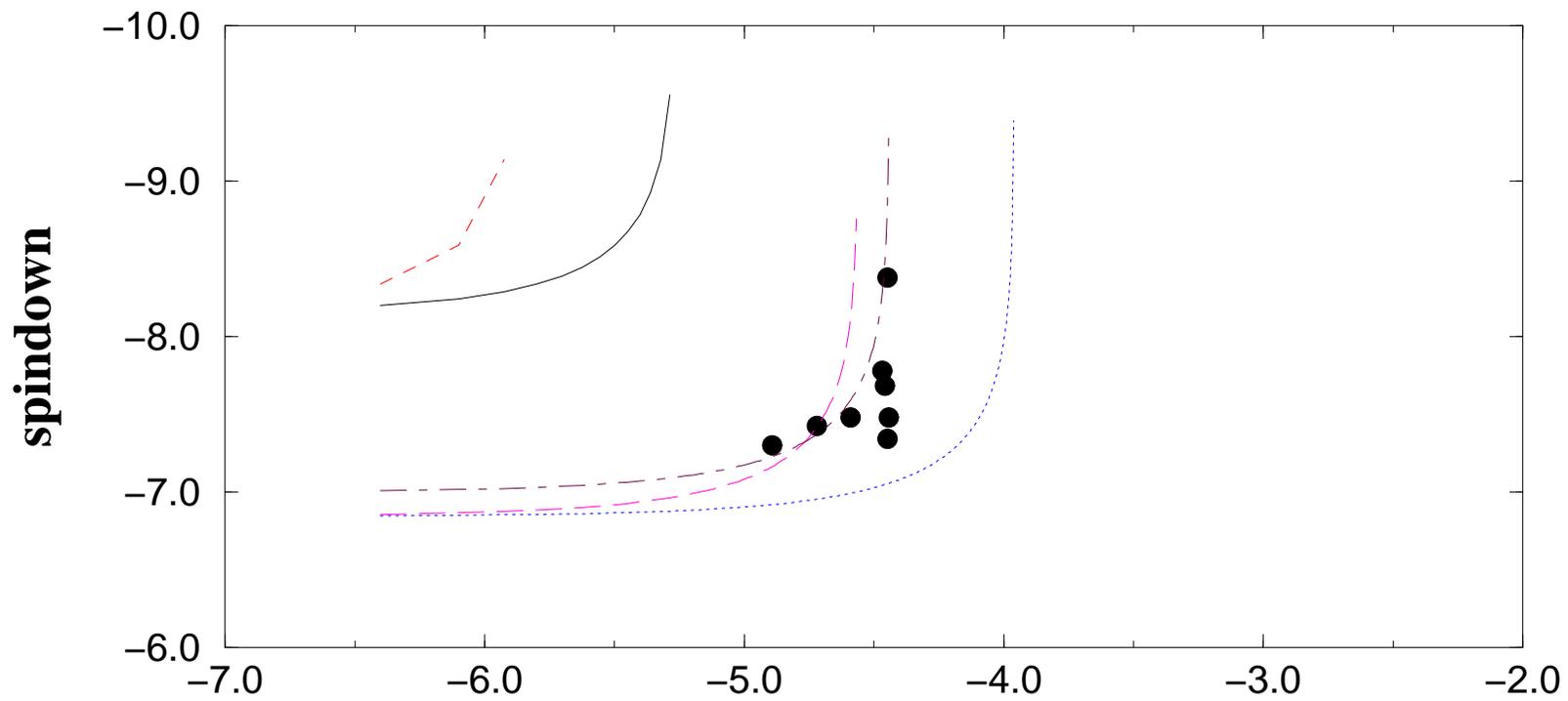
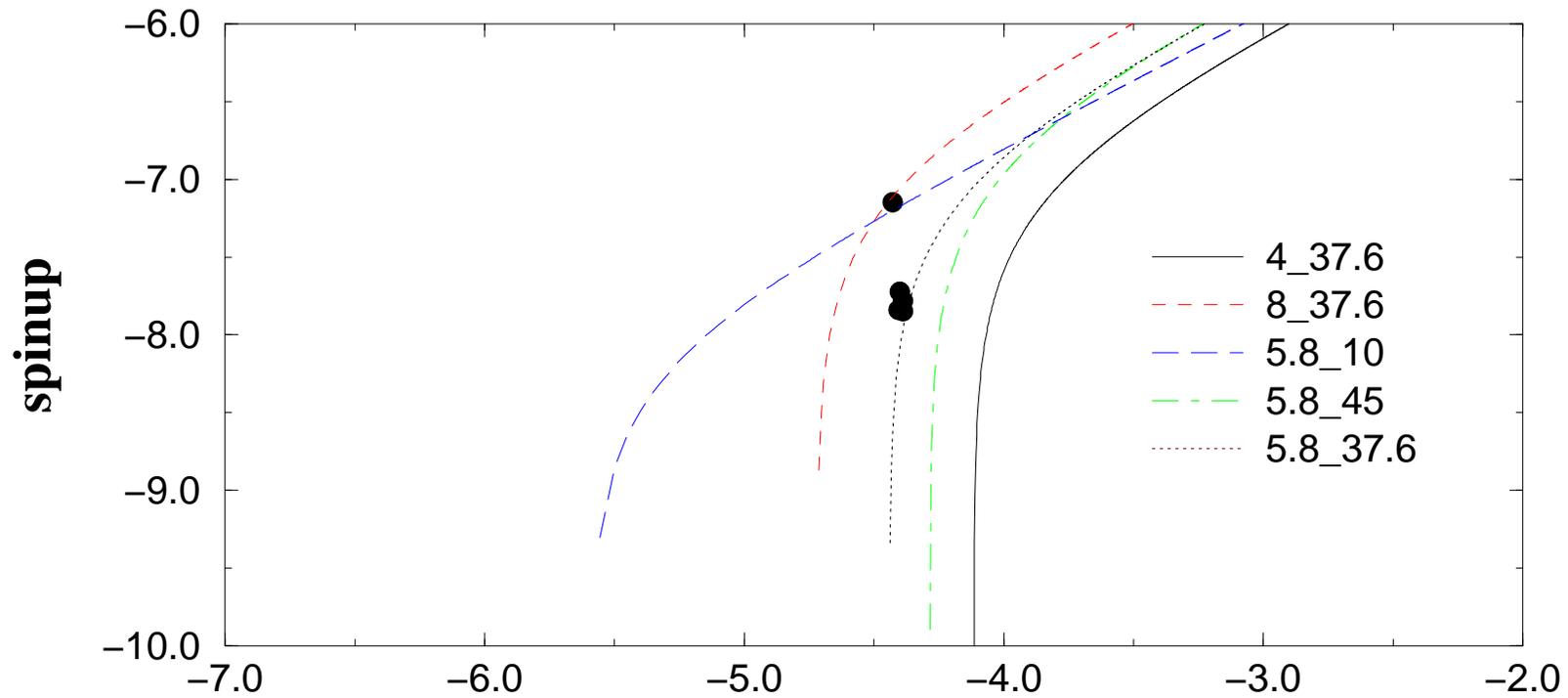
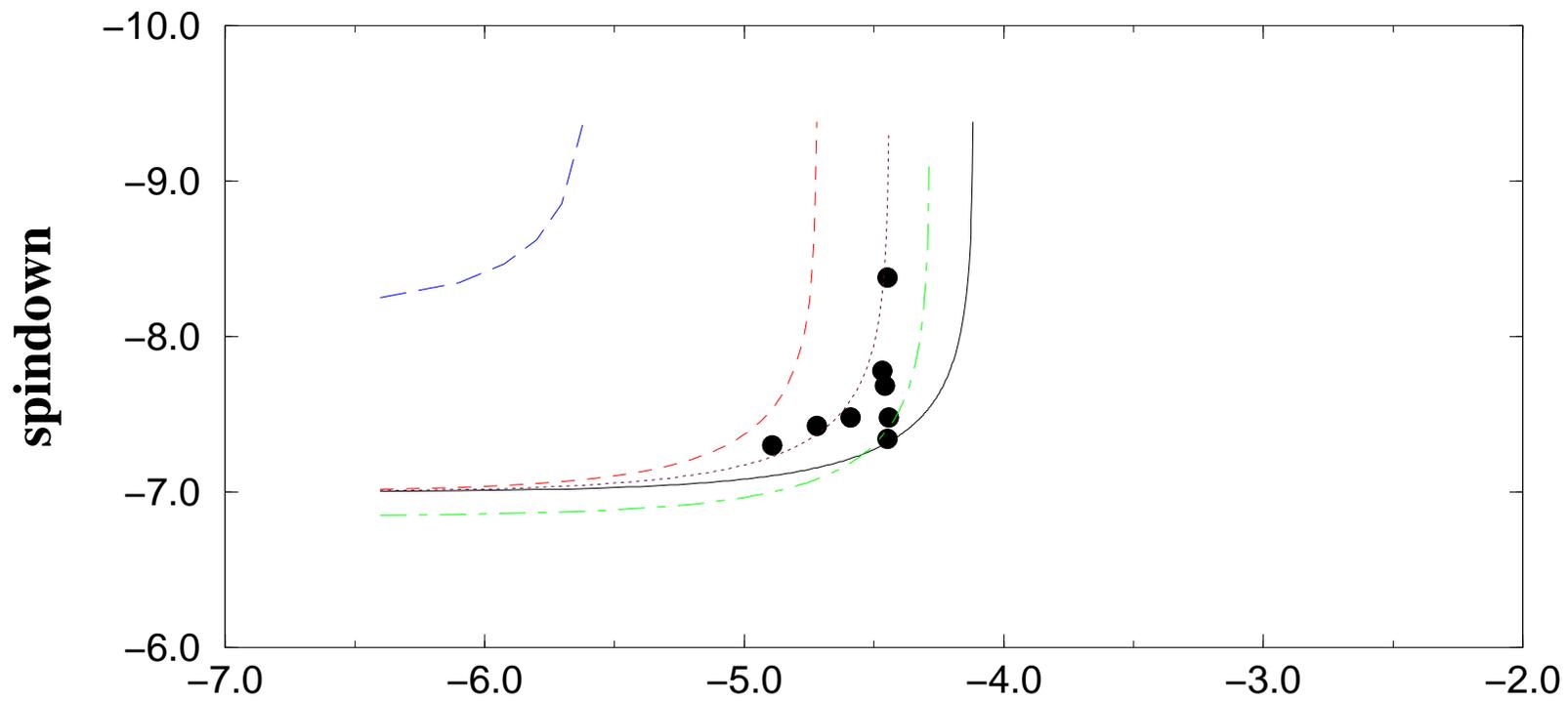
\end{document}